\documentclass[10pt,letterpaper,twocolumn]{article} 
\usepackage{ol2}

\usepackage[draft]{hyperref}
\usepackage{amsmath}

\usepackage{times} 

\usepackage{bm} 
\usepackage{graphicx,graphics} 

\begin{document}
\twocolumn[ 


\title{Fast fluorescence dynamics in non-ratiometric calcium indicators}

\author{M. Gersbach,$^1$ D. L. Boiko,$^{2,*}$ C. Niclass,$^1$ C. Petersen,$^1$ and E. Charbon$^1$}

\address{
$^1$Ecole Polytechnique F\'ed\'erale de Lausanne, Quantum
Architecture Group, 1015, Lausanne, Switzerland \\
$^2$Centre Suisse d'Electronique et de Microtechnique SA, 2002,
Neuch\^atel, Switzerland \\
$^*$Corresponding author: dmitri.boiko@csem.ch
}





%


\begin{abstract}
The fluorescence decay of high-affinity non-ratiometric Ca$^{2+}$
indicator Oregon Green BAPTA-1 (OGB-1) is analyzed with unprecedented
temporal resolution in two-photon excitation regime. A triple
exponential decay is shown to best fit the fluorescence dynamics
of OGB-1. We provide a 
model for accurate measurements of the
free Ca$^{2+}$ concentration and dissociation constants of
non-ratiometric calcium indicators.
\end{abstract}


 ] 

\maketitle

\newpage

Fluorescence Lifetime Imaging Microscopy (FLIM) is used to
locally probe the chemical environment of fluorophores, e.g., ion
concentration, pH, or oxygen content \cite{Lakowicz99,Suhling05}.
To acquire time-resolved fluorescence images, the technique of
Time-Correlated Single Photon Counting (TCSPC),
in combination with detectors exhibiting single-photon
sensitivity\cite{Cova83}, is commonly used. This technique
enables the measurement of photon time-of-arrival distributions
with very high accuracies, independently of instabilities in the
excitation beam intensity. So far, temporal resolutions of a few
hundred picoseconds were considered sufficient in bio-medical FLIM
applications.

In this letter, we analyze 
the fluorescence decay of the
high-affinity Ca$^{2+}$ indicator Oregon Green BAPTA-1 (OGB-1)
under two-photon excitation conditions, using a TCSPC system of
measured resolution 79ps based on CMOS Single Photon Avalanche Diode
(SPAD) detector technology.
Our measurements reveal a 
triple-exponential decay of OGB-1
fluorescence, which we show enables accurate measurements of
Ca$^{2+}$ 
concentration and dissociation constants of non-ratiometric
fluorescent probes. We provide 
a 
comparison
with previously reported data
\cite{Lakowicz99,Agronskaia04,Wilms06}, which were acquired
at resolution
of 200ps. 

\begin {figure}[htb]
\centerline{
\includegraphics{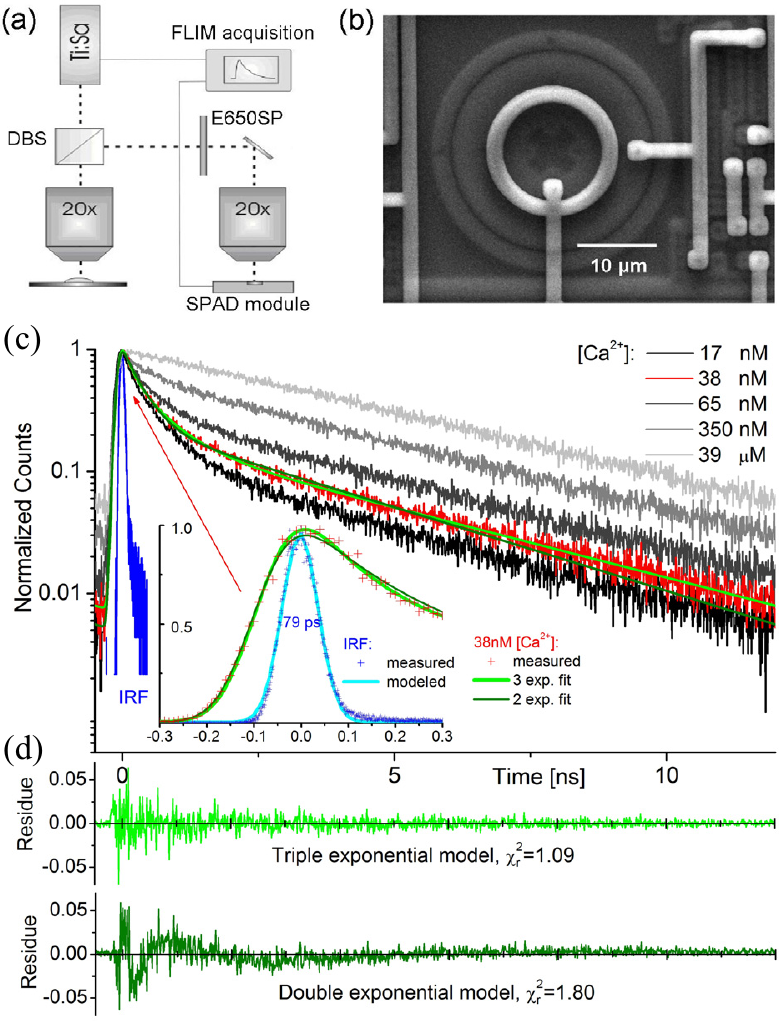}
} \caption {\label {FIG1} (Color online) (a) Schematic of the experimental setup
for fluorescence lifetime measurements. (b) Scanning electron
microscope image of a SPAD. (c) Measured fluorescence decay of
OGB-1 at 
various calcium concentrations (gray scale curves) and
hyper-Rayleigh scattering from colloidal gold particles
(blue curve). The red curve highlights measured
fluorescence decay in a 38 nM free Ca$^{2+}$ concentration buffer
and the numerical fits of the double-exponential (olive curve, $\chi^2_r$=1.80) and
triple-exponential (green curve, $\chi^2_r$=1.09) decay models. The inset shows a
close-up of the initial interval of 600ps width and (d) shows the
residues for the two models. }
\end{figure}

The SPAD-based TCSPC system is depicted in Fig.\ref{FIG1}. Our
SPADs \cite{Niclass05} are integrated in a 32x32 array and
incorporate on-chip high bandwidth I/O circuitry. The active
region of a SPAD pixel [Fig.\ref{FIG1}(b)] consists of a p$^+$-n
junction operating in the Geiger mode. Due to a small diameter (7
$\mu$m), our SPADs show an extremely low dark count rate
(${<} 10$ Hz at room temperature). The photon detection
probability is 25 $\%$ at 500 nm wavelength, while the dead time
is 25 ns with negligible afterpulsing ($<0.1~\%$)
\cite{Niclass05}.

Fluorescent molecules are excited in the two-photon regime
\cite{Denk90}, using a mode-locked Ti:Sapphire laser (MaiTai,
Spectra Physics) emitting 100 fs optical pulses at 800 nm
wavelength [Fig.\ref{FIG1}(a)]. The attenuated beam with an
average power of 9 mW is focused on a sample using a 20x
microscope objective (XLUMPlanFL, Olympus), which also serves to
collect the fluorescent emission. Fluorescence from OGB-1 samples
treated here (central emission wavelength 520 nm) is directed
towards the SPAD by a dichroic beam splitter (DBS) and a filter
(E650SP, Chroma Technology) for suppression of backscattered
excitation pulses. Another 20x objective images the emission spot
onto the SPAD. The low quantum yield of the two-photon excitation
fluorescence
results in a count rate at the detector of ~10 kHz for 80 MHz
repetition rate of excitation pulses. Time discrimination is
performed in a "reversed start-stop" configuration by measuring
the time intervals from a photo-detection event at the SPAD to
emission of a successive laser pulse. For precise timing of
photon arrivals, we use a 6 GHz bandwidth oscilloscope
(WaveMaster, LeCroy) incorporating a
time-to-digital converter 
and enabling computation of
histograms of photon arrivals.

Fig.\ref{FIG1}(c) shows the Instrument Response Function (IRF) of
the entire system at 400 nm wavelength (blue curve). It is
recorded by measuring the hyper-Rayleigh scattering of 800 nm
wavelength pulses in a solution of colloidal gold particles
(G1652, Sigma-Aldrich) \cite{Habenicht02}. At the incident power
of 90 mW, the average count rate of the detector is just 600 Hz.
The measured photon arrival time jitter of 79 ps (FWHM) is
dominated by the SPADs time-response characteristics. The IRF is
only slightly asymmetrical and assumes a Gaussian-curve
approximation $\textrm{IRF}{=}\exp
(-t^2/2\sigma_{\textrm{IRF}}^2)$ (inset, cyan curve).

Fluorescent samples were composed of $2~\mu$l OGB-1 dye (O6806,
Molecular probes) and $20~\mu$l Ca:EGTA buffer solutions from the
calibration kit (C3008MP, Molecular Probes) with quoted free
Ca$^{2+}$ concentrations in the range of 17 nM to 39 $\mu$M.
Several fluorescence decay curves measured in the two-photon
excitation regime are shown in Fig.\ref{FIG1}(c). The
fluorescence lifetimes were obtained from the numerical analysis
of these data.

Thorough numerical fit must take into account the IRF of the
system \cite{Wilms06}. As opposed to conventional systems with
strongly asymmetric IRF, the Gaussian-like IRF of our system
assumes a simple analytical expression for the measured
fluorescence decay. 
For a train of excitation pulses of period $T$, each term of a
multi-exponential decay process reads
\begin{equation}
\begin{split}
I_k&=\frac 12 \Bigl[\frac{1+e^{-T/\tau_k}}{1-e^{-T/\tau_k}}-\textrm{erf}\Bigl(\frac{\sigma_{\textrm{IRF}}}{ \sqrt2 \tau_k}- \frac{t}{\sqrt2 \sigma_{\textrm{IRF}}} \Bigl) \Bigr] \\
&\times
\exp\Bigl(-\frac{t}{\tau_k}+\frac{\sigma_{\textrm{IRF}}^2}{2\tau_k^2}
\Bigr), \qquad \qquad k=\{f,i,s\}
\end{split}
\label{EQ1}
\end{equation}
where a triple-exponential decay is assumed and index $k$
indicates fast ($f$), intermediate ($i$)  and slow
($s$)  temporal components, $\tau_k$ is the fluorescence emission
lifetime ($\tau_f{<}\tau_i{<}\tau_s$) and
$\textrm{erf}(z){=}\frac{2}{\sqrt{\pi}}\int_0^z \exp(-\xi^2)d\xi$
is the error function. Eq.(\ref{EQ1}) takes the periodic train of
excitation pulses and response time jitter of the SPAD into
account, such that our data in Fig.\ref{FIG1} do not require
deconvolution processing.

Analysis of the OGB-1 fluorescence reveals the best agreement
with a triple exponential decay approximation. The data are
modeled using the function $A_fI_f{+}A_iI_i{+}(1{-}A_f{-}A_i)I_s$
with the fast $I_f(t)$, intermediate  $I_i(t)$ and slow  $I_s(t)$
decaying components (\ref{EQ1}) of normalized partial intensities
$A_f$, $A_i$ and $A_s{=}1{-}A_f{-}A_i$, respectively.
Fig.\ref{FIG1}(c) details a comparison between the double-exponential (olive curve)
and triple-exponential (green curve) models applied to
fluorescence from a 38 nM Ca$^{2+}$ concentration sample (red
points and curve). In both cases, the residues [Fig.\ref{FIG1}(d)]
do not show any bias that might be caused by the Gaussian curve
approximation of the IRF, while the quality
of the numerical fit
is improved in case of the triple-exponential model, as also confirmed by the reduced $\chi^2$-values. Wilms et al
\cite{Wilms06} have made the same observation for OGB-1
fluorescence in the absence of Ca$^{2+}$. However, it was
attributed to contaminating dye derivatives and a
double-exponential model has been utilized for samples containing
Ca$^{2+}$.
Previously reported data \cite{Lakowicz99,Agronskaia04,Wilms06}
for OGB-1 fluorescence lifetimes thus assume double-exponential
decay. 
We argue that at high Ca$^{2+}$ concentration, large amplitude of
long-leaving component make it difficult to detect short lifetime
components with small amplitudes and the FLIM 
systems used in \cite{Lakowicz99,Agronskaia04,Wilms06}
exhibit insufficient temporal resolution to enable
observation of the triple-exponential behavior.

\begin{figure}[ftb]
\centerline{
\includegraphics{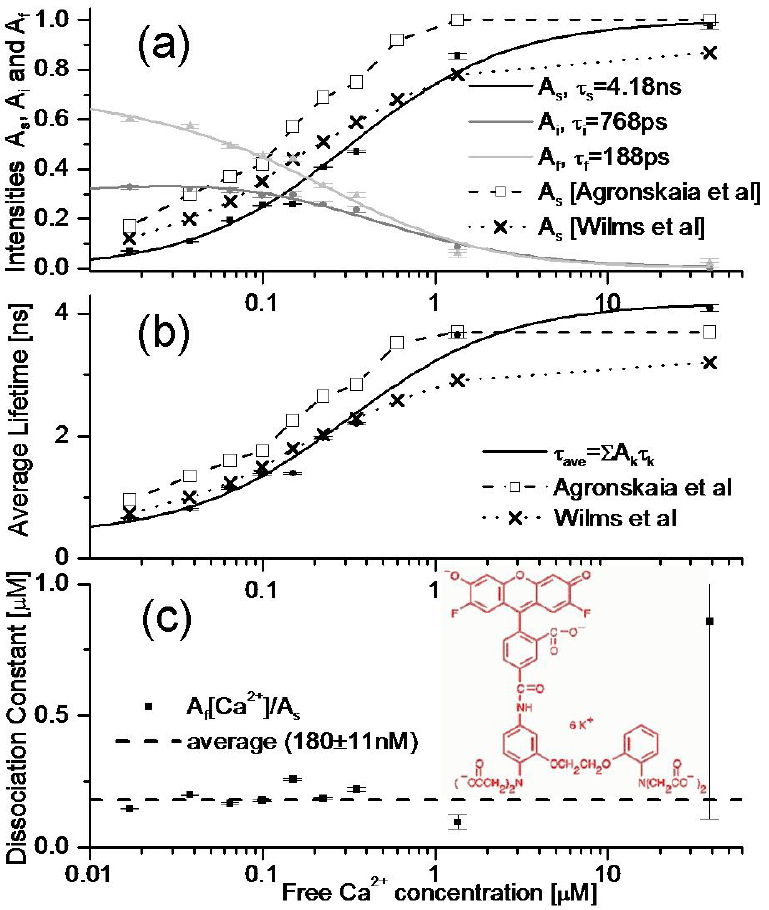}
}
\caption{(Color online) Partial intensities $A_s$, $A_i$ and $A_f$ of the slow ($\tau_s$=4.18${\pm}$0.01ns),
intermediate ($\tau_i$=768${\pm}$16ps) and fast ($\tau_f$=188${\pm}$6ps) decay components (a) and the mean lifetime $\tau_{\text{ave}}{=}{\sum_k} A_k \tau_k$ (b) of OGB-1
fluorescence as a function of the free Ca$^{2+}$ concentration in comparison with the data
from [\onlinecite{Agronskaia04}] and [\onlinecite{Wilms06}], $\chi_r^2$=0.96.
The numerical fit to the model (\ref{EQ2}) (curves) yields 
$K_D$=195${\pm}$9 nM, $K_{Dn}$=4${\pm}$6 $\mu$M and n=9${\pm}$3.
(c): The ratio $A_f[\textrm{Ca}^{2+}]/A_s$ (points) and
its variance-weighted mean yielding $K_{D}$=180${\pm}$11nM (dashed line).
The
inset shows the structure of OGB-1 molecule.} \label{FIG3}
\end{figure}

Using triple-exponential fluorescence decay (\ref{EQ1}) with the
lifetimes independent of the calcium buffer
\cite{Lakowicz99,Wilms06}, we applied a global numerical analysis
to our data, yielding the partial intensities $A_k$
in function of free Ca$^{2+}$ concentration and the lifetimes $\tau_k$
(Fig.\ref{FIG3}). Temporal resolution of our system allows the
short lifetime component ($\tau_f{\sim}188{\pm}6\textrm{ps}$) to be
unambiguously resolved in the background of the intermediate- and
long-living fluorescence ($\tau_i{\sim}768{\pm}16\textrm{ps}$ and
$\tau_s{\sim}4.18{\pm}0.01\textrm{ns}$). The partial intensities $A_f$ and
$A_i$ decrease while the slow-component intensity
$A_s{=}1{-}A_f{-}A_i$ increases with concentration [Ca$^{2+}$].

We attribute, as usual, the short- and long-lifetime components
to, respectively, unbound and Ca$^{2+}$-bound OGB-1 with a simple
1:1 complex stoichiometry. As such and because the intensities
are defined by complex concentrations, we obtain  that $A_f
{\propto} [\textrm{D}] $ and $A_s {\propto}
[\textrm{CaD}]{=}K_D^{-1} [\textrm{Ca}^{2+}][\textrm{D}]$ with
$[\textrm{D}]$ and $[\textrm{CaD}]$ being the unbound and
Ca-bound OGB-1 concentrations and $K_D$ the dissociation constant
of the 1:1 complex. The ratio $A_f[\textrm{Ca}^{2+}]/A_s$ thus
yields an estimate of the dissociation constant. In
Fig.~\ref{FIG3} (c), a weighted average ratio $180{\pm}11$nM is in good
agreement with the quoted (by manufacturer) $K_D$ of 170 nM.

In Fig.\ref{FIG3}, the long lifetime component $A_s$ is in good
agreement with the measurements of Agronskaia \textit{et al}\cite{Agronskaia04}
and Wilms \textit{et al}\cite{Wilms06}, but in [\onlinecite{Wilms06}], $A_s{<}1$
in Ca$^{2+}$-saturated buffer and $K_D$ is too high ($\sim$300
nM). These discrepancies are attributed to the dye impurities and
to the difference in two-photon absorption cross-sections of
Ca-bound and unbound OGB-1, which are said difficult to be
quantified \cite{Wilms06}.

For the Ca-bound OGB-1, Lakowicz\cite{Lakowicz99} reported
$\tau_s$ of 4 ns, which agrees well with our data. In
[\onlinecite{Agronskaia04}], $\tau_s$ varies in the range 2.6-3.7
ns, while in [\onlinecite{Wilms06}], $\tau_s{=}3.63$ ns. For the
Ca-free OGB-1, a single lifetime component $\tau_f$ has been
measured of 700 [\onlinecite{Lakowicz99}], 290-420
[\onlinecite{Agronskaia04}] and 346 ps [\onlinecite{Wilms06}],
respectively. These results correspond to the combined effect of
the decay processes with the lifetimes $\tau_f{\sim}190$ps and
$\tau_i{\sim}770$ps in our measurements, which may not have been
resolved in previous studies. The experimental setup in
[\onlinecite{Wilms06}] relies on a commercially available Photomultiplying Tube (PMT)
with a quoted timing jitter of 200 ps. Other
data\cite{Lakowicz99,Agronskaia04} are reported without timing
resolution of experimental setups.

Due to the Ca-binding features of the octadentate chelator BAPTA
\cite{Tsien80}, the 1:1 stoichiometry is usually attributed to the
Ca-bound OGB-1 
as well. However, the molecule of
OGB-1 has an asymmetric structure with respect to its BAPTA
moiety (Fig.\ref{FIG3}, inset). The asymmetric arrangement of the
carboxyl functional groups,  benzol rings, fluorine and nitrogen
atoms as well as the variations of electronic density from low
(at hydrogen in carboxyl groups) to high (at benzol rings, F and
O atoms) assume a polarity of the molecule and, as a consequence,
weak dipole-dipole intermolecular forces. As a result of such
interaction forces, several OGB-1 molecules might be coordinated
to a Ca-bound OGB-1 forming thus a polymolecular association with
a calcium:indicator molar ratio 1:n. The intermediate lifetime
component $\tau_i$ in Fig. \ref{FIG3} then might be attributed to
such polymolecular association, yielding $A_i{\propto}
[\textrm{Ca}_{1/n}\textrm{D}] {=} K_{Dn}^{-1/n} [\textrm{D}]
[\textrm{Ca}^{2+}]^{1/n}$ with $K_{Dn}$ being the corresponding
dissociation constant. Note that formation of a polymolecular
structure is a very complex processes and requires thorough
investigations but a non 1:1 stoichiometry of Ca:OGB-1 has been
reported in \cite{Thomas00} and here it allows us to build an
accurate model:
\begin{equation}
\frac{[\textrm{Ca}^{2+}]}{K_D}{=}\frac {A_s}{
A_f}{=}\frac{A_s}{1{-}A_s}\biggl[1{+}\frac {A_i}{ A_f} \biggr],
\quad \frac{[\textrm{Ca}^{2+}]}{K_{Dn}}{=}\biggl(\frac {A_i}{
A_f}\biggr)^{n}. \label{EQ2}
\end{equation}
In the limit $K_{Dn}{\rightarrow}\infty$ (double exponential
decay), it agrees with the Hill equation, as opposed to the model
in [\onlinecite{Wilms06}]. The numerical fit of data in
Fig.\ref{FIG3} reports small binding affinity ($K_{Dn}{\sim} 4
\mu$M) and $n$=9 indicating the most probable form of
polymolecular association with eight Ca-free molecules
coordinated to the Ca-bound OGB-1. The dissociation constant
$K_D$ reported by the fit is 195 nM, in agreement with the
average of $A_f[\textrm{Ca}^{2+}]/A_s$ (bottom panel).

The triple exponential fluorescence decay in non-ratiometric Ca
probes as a result of dye contaminations was suggested by
Lakowicz \textit{et al}\cite{Lakowicz92} for 
the Calcium Green (CG-1).
The  CG-1 molecule has the same structure as OGB-1 but the two F
atoms are replaced by Cl, yielding\cite{Lakowicz92}
$\tau_f{=}50$ps, $\tau_i{=}450$ps and $\tau_s{=}3.7$ns.
Interestingly, the signature of non-1:1 stoichiometry has been
noticed for Ca-bound CG as well \cite{Eberhard91}. We find that at
low Ca$^{2+}$ concentration, the ratio
$A_f[\textrm{Ca}^{2+}]/A_s$ in [\onlinecite{Lakowicz92}] is a
constant of ~170$\pm$20 nM, yielding $K_D$ close to the quoted
value of 190 nM. At high Ca concentration, the band pass of the
system (2GHz modulation using a frequency-domain FLIM technique)
was insufficient to accurately measure the small component $A_f$.
The data in [\onlinecite{Lakowicz92}] can thus be accurately
interpreted in the framework of our model (\ref{EQ2}), without
appealing to dye impurities. (
$K_D$ obtained  in \cite{Lakowicz92}
using a conventional model\cite{Grynkiewicz85} is 128nM).

In summary, we have shown that high-temporal resolution
measurements of the triple-exponential fluorescence decay of
non-ratiometric Ca$^{2+}$ indicators allow the free Ca$^{2+}$
concentration and dye dissociation constant to be measured
precisely.

DLB is grateful to Leonid Zekel and Edwin Constable for
discussions on the model. This research was supported, in part,
by a grant of the Swiss National Science Foundation and by Centre
SI of EPFL.

\end{document}